\documentstyle[epsf]{article}%
\begin{document}

\title{Prediction of protein secondary structure based on residue pairs}
\author{Xin Liu${^1}$,Li-Mei Zhang${^2}$,
Wei-Mou Zheng${^3}$\\\\\\\\\\\\\\\\\
\small
${^1}${\it The Interdisciplinary Center of Theoretical Studies, Chinese Academy of Sciences, Beijing 100080, China}\\
\small${^2}${\it School of Science at Beijing Jiaotong University,
Beijing 100044, China} \\
\small${^3}${\it Institute of Theoretical Physics, China, Beijing
100080, China}}

\maketitle

\newpage
\begin{abstract}
The GOR program for predicting protein secondary structure is
extended to include triple correlation. A score system for a
residue pair to be at certain conformation state is derived from
the conditional weight matrix describing amino acid frequencies at
each position of a window flanking the pair under the condition for
the pair to be at the fixed state. A program using this score system
to predict protein secondary structure is established. After
training the model with a learning set created from PDB\_SELECT,
the program is tested with two test sets. As a method
using single sequence for predicting secondary structures, the
approach achieves a high accuracy near 70\%.
\end{abstract}

\leftline{PACS number(s): 87.10.+e,02.50.-r}%
\bigskip

\section{Introduction}

Methods for predicting the secondary structure of a protein from
its amino acid sequence have been developed for 3 decades. Besides
neural network models and nearest-neighbor methods, the
statistically based Chou-Fasman/GOR method is well-established and
commonly used. In 1974, assuming an oversimplified independency to
cope with the large size 20 of the amino acid alphabets at a small
size of database, Chou and Fasman (1974) derived a table of
propensity for a particular residue to be in a given secondary
structure state. By combining with a set of rules, the protein
secondary structure was predicted using this propensity. Later, in
the first version of the GOR program (Garnier, Osguthorpe, and
Robson, 1978), the state of a single residue $a_i$ was predicted
according to a window from $i-8$ to $i+8$ surrounding the residue.
Unlike Chou-Fasman which assumes that each amino acid individually
influences its own secondary structure state, GOR takes into
account the influence of the amino acids flanking the central residue
on the central residue state by deriving an information
score from the weight matrix describing 17 individual amino acid
frequencies at sites $i+k$ with $-8\leq k\leq +8$. By using a
single weight matrix, the correlation among amino acids within the
window was still ignored. In the later version GOR III (Gibrat,
Garnier, and Robson, 1987), instead of single weight matrix for
every structure state, 20 
weight matrices, each of which corresponds to a specific type of
the central residue, were used. These conditional weight matrices
take the pair correlation between the central residue and a
flanking one into account. In the most recent version of GOR (GOR
IV, Garnier, Gibrat, and Robson, 1996), all pairwise combinations
of amino acids in the flanking region were included.

The GOR program maps a local window of residues in the sequence to
the structural state of the central residue in the window.
Correlations among positions within the window is essential for
improvements in prediction accuracy. We give an
example to show the importance of high order correlations. For
central residue $a_i=K$ being at the extended strand state, the
conditional probability for $a_{i-3}=V$ is $P^e_{-3;0}(V|K)
=0.088$, while the conditional probability for $a_{i-3}=V$ at
$a_i=K$ and $a_{i+1}=E$ is $P^e_{-3;0,+1}(V|KE) =0.186$, and
$P^e_{-3;0,+1}(V|KV) =0.058$. With the growth of protein
structure database, now the size of known protein structures
allows us to consider correlations higher than pair ones.
Here we shall extend the GOR program
to include triple correlations, developing a program to predict
protein secondary structure based on residue pairs (PSSRP). As a
method using single sequence, the computation required is rather
light, but its prediction accuracy reaches 70\%.

\section{Methods}

Kabsch and Sander (1983) define eight states of secondary
structure according to the hydrogen-bond pattern. As in most
methods, we consider 3 states $\{h, e, c\}$ generated from the 8
by the coarse-graining $H,G,I\to h$, $E\to e$ and $X,T,S,B\to c$.

\subsection{Window-based scores}
A window is a sequence segment $a_ia_{i+1}\ldots a_{i+l-1}$ of the
width $l$. Consider two residues $a_{i+j}=x$ and $a_{i+k}=y$
inside the window (with $0< j< k< l-1$). Their conformation state
are $\alpha$ and $\beta$, respectively. Let us denote by $w$ the set
of the sites within the window with $i+j$ and $i+k$ excluded. When
discussing probability, we ignore the starting site index $i$.

The Chou-Fasman propensity of residue $x$ to conformation $\alpha$
is defined by
\begin{equation}
CF(x;\alpha) =\frac{P(x|\alpha )}{P(x)},
\end{equation}
where $P(x)$ is the probability for residue $x$ to appear, and
$P(x|\alpha )$ the conditional probability for $x$ to be at
conformation $\alpha$. Here we use only a logarithmic propensity,
the logarithm of $CF$:
\begin{equation}
LCF(x;\alpha)=\log CF(x;\alpha).\end{equation}
As an extension of the Chou-Fasman propensity
of residues, the propensity of residue pair $xy$ to conformation
$\alpha\beta$ may be defined as
\begin{equation}
r_d(xy;\alpha\beta)=\frac {P_d(xy|\alpha\beta )}{P_d(xy)},\quad
d=k-j-1,
\end{equation}
where $P_d(xy)$ is the probability for residue pair $x$ and $y$ to
appear with their site index difference being $d$, and
$P_d(xy|\alpha\beta )$ the conditional probability on the
condition that the conformation states of $x$ and $y$ are $\alpha$
and $\beta$, respectively.

When a window flanking the pair $xy$ is examined to infer the
conformation $\alpha\beta$ of $xy$, a further extension of the
Chou-Fasman propensity is
\begin{equation}
R_d(xy;\alpha\beta)=\frac {P_d(xy,w|\alpha\beta
)}{P_d(xy,w)},\quad d=k-j-1,
\end{equation}
where $P_d(xy,w)$ and $P_d(xy,w|\alpha\beta )$ are now probability
for the whole window, i.e. $xy$ and $w$. Making the assumption of
independency, we introduce the conditional weight matrix ${\bf Q}_{xy}$
of $(l-2)$ columns and 20 rows, whose entries describe the probability
for a specific residue $z$ to appear at some flanking site, say the $n$-th
site from the window starting position. We denote the probability by
$Q_{d,n}(z|xy)$, and write
\begin{equation}
P_d(xy,w)=P_d(xy)P_d(w|xy)=P_d(xy)\prod_{n\in w} Q_{d,n}(a_{i+n}|xy).
\end{equation}
A similar simplification for $P_d(xy,w|\alpha\beta )$ is
\begin{equation}
P_d(xy,w|\alpha\beta )=P_d(xy|\alpha\beta ) P_d(w|xy,\alpha\beta ) =
P_d(xy|\alpha\beta )\prod_{n\in w} Q_{d,n}(a_{i+n}|xy,\alpha\beta ),
\end{equation}
where the meaning of $Q_{d,n}(z|xy,\alpha\beta )$ is analogous to
$Q_{d,n}(z|xy)$. The window score $I_d(xy;\alpha\beta )$ for pair
$xy$ to be at conformation $\alpha\beta$ is then defined as the
logarithmic ratio of $R_d$:
\begin{equation}
I_d(xy;\alpha\beta )= \log\left[\frac {P_d(xy|\alpha\beta )}
{P_d(xy)}\right] +\sum_{n\in w} \log\left[ \frac{Q_{d,n}
(a_{i+n}|xy,\alpha\beta )}{Q_{d,n}(a_{i+n}|xy)} \right].
\end{equation}

So far we have not determined the window width $l$ and the
position of the inferred residue pair inside the window,
especially the separation $d$ of the two residues. To do this, we
need a measure of the distance between two probability
distributions. A well defined measure is the Kullback-Leibler (KL)
distance or relative entropy (Kullback et al., 1959; Kullback, 1987;
Sakamoto et al., 1986), which, for two distributions
$\{p_i\}$ and $\{q_i\}$, is given by
\begin{equation}
KL(\{p_i\}, \{q_i\})= \sum_i p_i \log(p_i/q_i).
\end{equation}
It corresponds a likelihood ratio, and, if $p_i$ is expanded around
$q_i$, its leading term is the $\chi^2$ distance. It is often to use
the following symmetrized form
\begin{equation}
D(\{p_i\}, \{q_i\})= \hbox{$\frac 12$}[KL(\{p_i\}, \{q_i\})+
KL(\{q_i\}, \{p_i\})].
\end{equation}
The distance $D_n(xy,\alpha\beta ) \equiv
D[\{Q_{d,n}(z|xy,\alpha\beta ) \}_z, \{Q_{d,n}(z|xy)\}_z ]$
measures the power for site $i+n$ to infer conformation
$\alpha\beta$. Asymptotically $D_n(xy,\alpha\beta )$ approaches
zero, when $n$ becomes far away from the sites of pair $xy$. The
power for a window to infer conformation $\alpha\beta$ of $xy$ may be
measured by $\left\langle \sum_{n\in w} D_n(xy,\alpha\beta
)\right\rangle$, the window sum distance averaged with the weight
$P(xy,\alpha\beta )$. We find a reasonable choice is $l=16$ and
$d=0,1$ with residue $y$ being at 9-th site of the window.
Detailed discussion will be published elsewhere. Due to limited
samples, we consider only $\alpha\beta \in \{cc, ee, hh, ce, ch,
ec, hc\}$ with $eh$ and $he$ excluded.

\subsection{Prediction steps}
Using scores $I_d(xy;\alpha\beta )$ and sliding windows, we may
calculate scores of true and false windows for each combination of
$xy;\alpha\beta$ and $d$. The threshold $T_d(xy;\alpha\beta )$ is
determined by the error rate 5\% at which non-$\alpha\beta$
conformations are wrongly predicted as $\alpha\beta$. For a
given window, if its score $I_d(xy;\alpha\beta )$ is greater than
the corresponding threshold $T_d(xy;\alpha\beta )$, we say that
`$x$ at $\alpha$', `$y$ at $\beta$' and `$xy$ at $\alpha\beta$'
are {\it evidenced} by the event $(d;xy;\alpha\beta )$.

\noindent{\bf Step 1:} If `$xy$ at $\alpha\beta$' is evidenced by
event $(d;xy;\alpha\beta )$, and, at the same time, both `$x$ at
$\alpha$' and `$y$ at $\beta$' are further evidenced by some other
events, we say that `$xy$ at $\alpha\beta$' is {\it strongly
confirmed}. Determine all strongly confirmed pairs.

\noindent{\bf Step 2:}

We now consider the case that `$xy$ at $\alpha\beta$' is
evidenced,  but not strongly confirmed. If either `$zx$ at
$\gamma\alpha$' or `$yz$ at $\beta\gamma$' is strongly confirmed,
we say that `$xy$ at $\alpha\beta$' is {\it weakly confirmed}.
Determine all the weakly confirmed pairs.

\noindent{\bf Step 3:}

From a confirmed pair, either strongly or weakly, we calculate the
score $I(x;\alpha )$ for single residue $x$ to be at conformation
$\alpha$ according to
\begin{equation}
I(x;\alpha )=\max_{d,d',y,z,\beta ,\gamma } \{ I_d(xy;\alpha\beta
)-LCF(y;\beta ), I_d'(zx;\gamma\alpha ) -LCF(z;\gamma )\}, \label{x}
\end{equation}
where only confirmed pairs are searched for maximum. The
conformation of $x$ is finally inferred as
\begin{equation}
\alpha^* ={\rm arg}_\alpha\max \{I(x;\alpha )\},
\end{equation}
i.e. $\alpha^*$ is the $\alpha$ corresponding to the maximal
$I(x;\alpha )$. Infer residue conformation for all the confirmed
pairs.

\noindent{\bf Step 4:}

In this step we expand already inferred $h$ and $e$ segments in
both direction. Suppose residue $x$ be a candidate for elongating
$e$. We calculate score $I(x;e)$ according to (\ref{x}), but now
only pairs fit `$x$ at $e$' are searched for maximum. If $I(x;e)$
is positive, we assign conformation $e$ to $x$. The elongation of
$h$ is similar.

\noindent{\bf Step 5:}

At both ends of the sequence no full windows are available. The
first and last two residues are always assigned to $c$. With
the contribution of the missing sites set as zero, scores
$I(x;\alpha )$ of some residue $x$ in the end regions are
calculated for the elongation of the already determined
boundary conformation $\alpha$ in a similar way to the last step.
For remaining residues we examine only the cases of four successive
residues, say $a_ia_{i+1}a_{i+2}a_{i+3}$, in the same confirmation
$\alpha$ by calculating $I_0(a_ia_{i+1};\alpha\alpha)+
I_0(a_{i+2}a_{i+3};\alpha\alpha)$, and then infer the conformation
according to the largest positive score.

\noindent{\bf Step 6:}

This final step is filtering. Each single residue $h$ and $e$
segment is discarded. A conformation segment $hh$ is expanded to
$hhh$ according to whichever neighbor site has a large score for
$h$. Each residue whose conformation cannot be predicted so far is
assigned to conformation $c$.

Let us explain the prediction steps in more words by a simple example.
Suppose that total of five events are found for segment {\tt PDEFGHI}
of a sequence as shown as follows.
\newpage
\begin{verbatim}
  ...ACLMNPDEFGHIKQRST...
          --e.c--
          e.e----
          ----c.h
          ----ch-
          ----cc-
\end{verbatim}
In step 1, {\tt EG} at {\tt ec} is strongly confirmed by the first
three events. This is the only strongly confirmed pair in the
segment. Based on the pair, in Step 2 we find that, all the
remaining four events are weakly confirmed. In Step 3, we can
easily infer `{\tt P} at {\tt e}', `{\tt E} at {\tt e}', `{\tt G}
at {\tt c}', and `{\tt I} at {\tt h}'. Further calculation of
$I(H;h)$ and $I(H,c)$ leads to `{\tt H} at {\tt h}'. The
conformation of {\tt PDEFGHI} is now inferred as {\tt e-e-chh}.
Step 4 then fills up the gaps to get {\tt eeeechh}.

\section{Result}

We create a nonredundant set of 1612 non-membrane proteins for
training parameters from PDB\_SELECT (Hobohm and Sander, 1994) with
amino acid identity less than 25\% issued on 25 September of 2001.
The secondary structure for these sequences are taken from DSSP
database (Kabsch and Sander, 1983). As mentioned above, the eight
states of DSSP are coarse-grained into 3 states: $h$, $e$ and $c$.
This learning set contains 268031 residues with known
conformations, among which 94415 are $h$, 56510 are $e$, and 117106
are $c$. The size of the learning set is reasonable for training
our parameters.

To convert observed amino acid counts into frequencies or
probabilities for scoring is a basic problem faced in training. A
practical approach is to use pseudocounts (Aitchison and Dumsmore,
1972). We estimate background amino acid frequencies $\{\rho_x\}$
directly from counts of the whole learning set. Then, we estimate
the weight matrix element $Q_{d,n}(z|xy,\alpha\beta )$ from the
count $N_{d,n}(z|xy,\alpha\beta )$ of amino acid $z$ at position
$n$ under the condition that residue pair $xy$ with separation $d$
is in conformation $\alpha\beta$ as follows.
\begin{equation}
Q_{n,z}=\frac{N_{n,z}+\sqrt{N_n}\rho_z} {N_n+\sqrt{N_n}},  \quad
N_n =\sum_z N_{n,z},
\end{equation}
where $Q_{n,z}$, indicating specifically only $n$ and $z=a_{i+n}$,
stands for $Q_{d,n}(a_{i+n}|xy,\alpha\beta )$, and $N_{n,z}$ stands for
$N_{d,n}(a_{i+n}|xy,\alpha\beta )$. Here, the conditional probability
$Q_{n,z}$ is estimated by using a pseudocount propotional to the
background distribution $\rho_z$. If $\sum_z N_{d,n}
(z|xy,\alpha\beta )$ is less than 10, the sample size is too small to
reliably estimate the score. No such scores will be used for inference.
Or, equivalently speaking, they are set to be negative infinity.

In order to assess the accuracy of our approach, we use the following 2
test sets: Sets 1 and 2. A set of 124 nonhomologous proteins is
created from the representative database of Rost and Sander (1993)
by removing subunits A and B of hemagglutinin 3hmg, which are
designated as membrane protein by SCOP (Murzin et al, 1995). The
124 sequences and the learning set are not independent of each
other according to HSSP database (Dodge, Schneider and Sander,
1998). That is, some proteins of the 124 sequences and certain proteins
in the learning set belong to the same putative homologue family
of HSSP. Removing these proteins from the 124 sequences and 5 seuqences
with unknown amino acid segments longer than 6, we construct Set 1 of 76
proteins, a subset of the 124 sequences. 34 proteins
with known structures of the CASP4 database issued in December of
2000 are taken as Set 2.

The predicted counts of each conformation type in the test sets
are  listed in Table 1. The quantities assessing accuracy on
single residue level for a given test set
are the total percent correct $Q_3$, percent of types $h$ and $e$
predicted correctly (sensitivity $S_n$) and percent of predictions
correct for types $h$ and $e$ (specificity $S_p$). Results obtained
on the test sets are listed in Table 2 in comparison with the results
of GOR IV and SSP (Solovyev and Salamov, 1991, 1994), another secondary
structure predictor based on discriminant analysis using single sequence.
The approach PSSRP performs very well. The overall value of $Q_3$
averaged over Sets 1 and 2 is 70.2\%. We show $Q_3$ statistics for
Sets 1 and 2 in Fig.~1.

Generally, strongly confirmed pairs are the portion with a high
confidence in prediction. Strongly confirmed pairs cover 58.4\%
and 53.8\% of Set 1 and 2, respectively. There is a clear
correlation between the coverage rate or the percentage of the
strongly confirmed pairs in the total length of a sequence and
accuracy $Q_3$, which are
 79.5\% and 77.4\% for the coverage portion, respectively.
To examine the correlation between the coverage rate $C$ of
strongly confirmed pairs and the whole sequence accuracy $Q_3$, we
conduct simple linear regression of $Q_3$ on $C$ for Sets 1 and 2,
as shown in Fig.~2 with regression line $Q_3=0.728 C+0.306$.
Correlation coefficient $r$ and standard deviation $\sigma$ are
0.779 and 0.0276 for Set 1, while for Set 2, they are 0.830 and
0.0209, and for the two sets total are 0.795 and 0.0253. Thus, the
coverage rate of the strongly confirmed pairs provides us a
self-checking confidence level of the prediction accuracy.

$Q_3$ measure gives an overall number of residues predicted
correctly.  It is well known that single-residue accuracy
sometimes poorly reflects the quality of prediction. Measures
concentrating on secondary structure segment prediction accuracy
would better reflect the nature of structure. A simple segment
overlap measure is: a segment is considered correctly predicted if
the predicted and observed segments have at least two amino acids
in common (Taylor, 1984). Results of this segment prediction
accuracy from our approach are listed in Table 3.

\section{Discussions}

We have presented an improved approach using single protein
sequence  to predict secondary structure. The improvement is
achieved by including triple correlations. Most recent
improvements in accuracy come from methods which are capable to
consider correlations nonlocal in sequence. Combining evolutionary
information via multiple alignments of homologous sequences is a
main way to include such correlations. Although methods using
single sequence generally cannot cope with nonlocal correlations
easily, their simple nature of requiring least computation in the
prediction step is still attractive.

There are rooms for further improvement of PSSRP. The original
scores are obtained for residue pairs. The scores are then converted to
those for single residues in the prediction Step 3. It is possible
to construct a scoring system to directly use pair scores by
introducing appropriate weighting. We may tune the thresholds to
compromise $S_p$ with $S_n$. We may integrate the segment length
statistics into the approach, e.g. by dynamic programming.

The size of the amino acid alphabets is 20. Number of parameters
increases drastically with the order of correlations considered. A
statistical model containing a tremendous number of parameters
will require a huge learning set to train parameters. Furthermore,
an over-complicated model can easily result in overfitting. It
seems that the overfitting is not too serious. A way to reduce the
number of parameters is to coarse-grain the 20 amino acids into a
small number of categories. This is under study.

\begin{quotation}
{We thank Dr. Shan Guan and Prof. Jing-Chu Luo for their kindly
help to our work.

This work was supported in part by the Special Funds for Major
National Basic Research Projects and the National Natural Science
Foundation of China.}
\end{quotation}

\newpage
Table 1. Predicted counts for each conformation type.\\
\begin{tabular}{|c|c|ccc|ccc|}\hline\hline
\multicolumn{2}{|c|}{}&\multicolumn{6}{c|}{Predicted}\\
\cline{3-8}
\multicolumn{2}{|c|}{}&\multicolumn{3}{c|}{Set 1}&\multicolumn{3}{c|}{Set 2}\\
\multicolumn{2}{|c|}{}&$h$&$e$&$c$ &$h$&$e$&$c$\\
\hline
&$h$           &3296&275 &1097&2097&205 &806\\
{Observed}&$e$ &271 &1815&912 &188 &1030&558\\
{}&$c$         &749 &765 &4999&504 &383 &2553\\\hline
\end{tabular}\\

Table 2. Single residue accuracies.\\
\begin{tabular}{|l c c c c c
c|}\hline\hline
   &Set& $Q_3$ & $S_n^h$ & $S_p^h$ & $S_n^e$ & $S_p^e$ \\ \hline
  PSSRP  & 1& 71.3 & 70.6 & 76.4 & 60.5 & 63.6 \\
  GOR IV & 1& 66.0 & 63.3 & 68.5 & 54.7 & 55.3 \\
  SSP    & 1& 66.8 & 68.1 & 69.0 & 55.3 & 60.0 \\ \hline
  PSSRP  & 2& 68.2 & 67.5 & 75.2 & 58.0 & 63.7 \\
  GOR IV & 2& 63.2 & 67.1 & 64.3 & 43.0 & 54.6 \\
  SSP    & 2& 61.2 & 66.3 & 63.3 & 45.7 & 55.6 \\
\hline
\end{tabular}\\

Table 3. A comparison of segment prediction accuracy for short and long
helices and sheets. Here a simple measure of segment overlap is used: a
predicted segment is couted as a true positive (TP) if the predicted
segment and an observed segments have at least two residues in commen.
With this definition the sensitivity $S_n$ and specificity $S_p$ are
calculated. \\
\begin{tabular}{|l|cc|cc|cc|cc|cc|cc|}\hline\hline
{}& \multicolumn{6}{c|}{ Set 1 } & \multicolumn{6}{c|}{Set 2}
\\ \cline{2-13}
   {}& \multicolumn{2}{c|}{GOR IV} &\multicolumn{2}{c|}{SSP} & \multicolumn{2}{c|}{PSSRP}
     & \multicolumn{2}{c|}{GOR IV} &\multicolumn{2}{c|}{SSP} & \multicolumn{2}{c|}{PSSRP}\\
    &                      $S_n$& $S_p$& $S_n$&$S_p$ &$S_n$ &$S_p$    &$S_n$&$S_p$& $S_n$& $S_p$&$S_n$&$S_p$\\\hline
  All helices            & 62.7 & 76.1 & 64.9 & 83.2 & 76.7 & 76.2    &63.2 &66.9 & 66.0 & 76.2 &71.8 &72.4\\
  Long helices($l>8$)    & 89.7 & 91.0 & 94.2 & 86.7 & 96.3 & 94.1    &84.4 &81.7 & 85.6 & 78.0 &93.1 &96.3\\
  Short helices($l\leq8$)& 54.7 & 55.3 & 38.9 & 56.5 & 59.3 & 70.0    &42.3 &48.4 & 46.6 & 56.5 &51.0 &64.2\\
  All sheets             & 65.5 & 59.5 & 57.0 & 74.6 & 71.4 & 66.2    &50.9 &60.0 & 49.3 & 73.9 &64.6 &67.4\\
  Long sheets($l>6$)     & 84.7 & 75.0 & 82.0 & 80.6 & 86.0 & 88.2    &64.4 &75.7 & 58.9 & 80.6 &86.3 &86.4\\
  Short sheets($l\leq6$) & 59.1 & 56.9 & 48.5 & 67.3 & 66.5 & 63.1    &47.7 &57.9 & 47.1 & 66.7 &59.4 &63.2\\ \hline
\end{tabular}\\

\newpage

\begin{figure}
\centerline{\epsfxsize=11cm \epsfbox{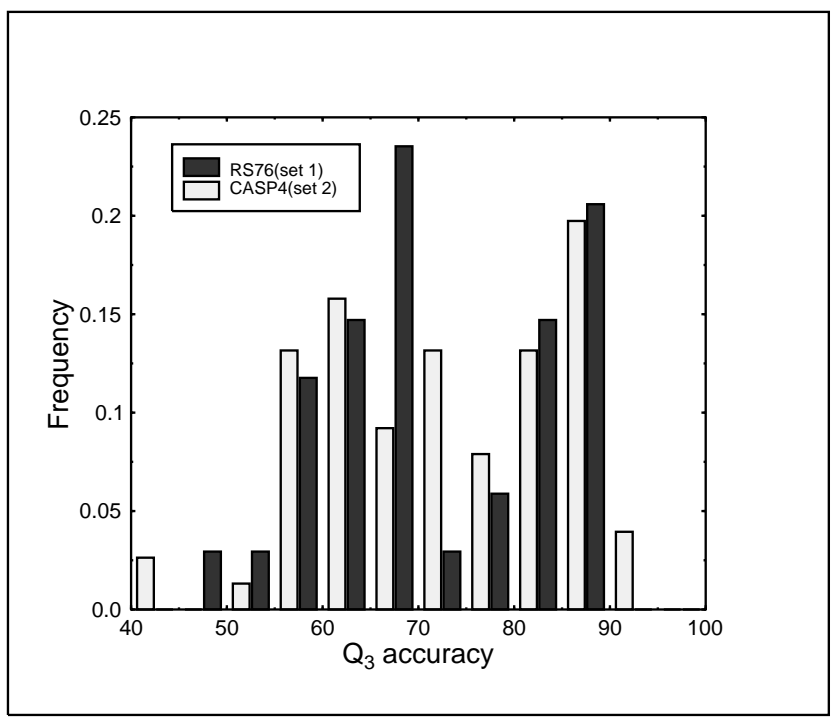}} \caption{$Q_3$
statistics for Sets 1 and 2. $Q_3$ is calculated for each
sequence, and the size of the bins is 0.05. } \label{fig1}
\end{figure}

\begin{figure}
\centerline{\epsfxsize=11cm \epsfbox{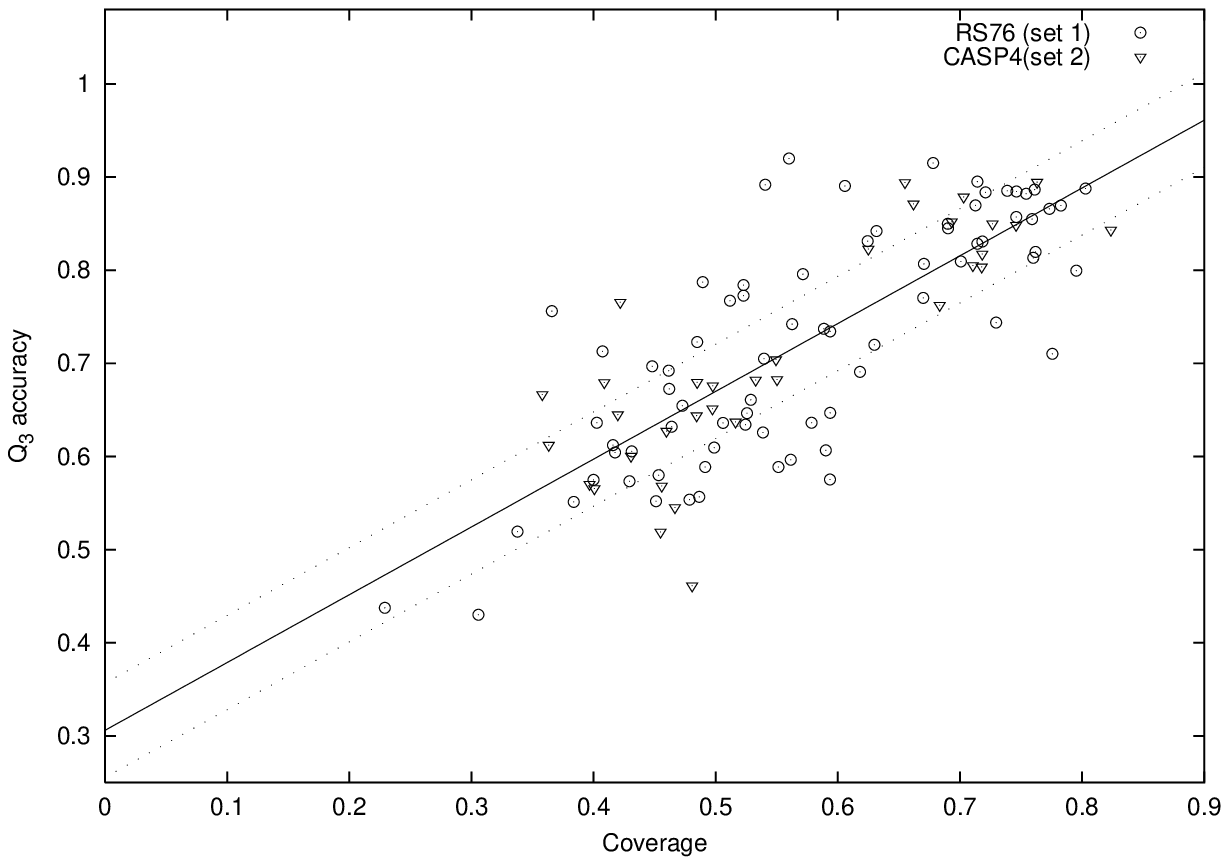}}\caption{Correlation
between whole sequence prediction accuracy $Q_3$ and the coverage
rate or the percentage of the strongly confirmed pairs in the
total length of a sequence.} \label{fig2}
\end{figure}


\begin{thebibliography}{99}
 {\parskip=0pt \parindent=0pt 
\bibitem{psdc} Aitchison,J. and Dumsmore,I.R. 1972. Statistical Prediction Analysis. Cambridge Univ. press, New York.
\bibitem{casp4} CASP4, http://predictioncenter.llnl.gov/casp4/
\bibitem{cf} Chou,P.Y.,and Fasman,G.D.1974(a).Conformational parameters for amino acids in helical,betasheet,and random coil regions calculated from proteins. Biochemistry 13(2),211-222.
\bibitem{cf1} Chou,P.Y.,and Fasman,G.D.1974(b).Prediction of protein conformation.Biochemistry, 13(2),222-245.
\bibitem{hssp} Dodge,C., Schneider,R., and Sander,C.1998.The HSSP database of protein structure-sequence alignments and family profiles. Nucleic Acids
Res, 26, 313-315.
\bibitem{gor4} Garnier,J.,Gibrat,J.F.,and Robson,B.1996.GOR method for predicting protein secondary structure from amino acid sequence.Methods in Enzymology 266,540-553.
\bibitem{gor} Garnier,J., Osguthorpe,D.,and Robson,B.1978.Analysis of the accuracy and implications of simple methods for predicting the secondary structure of globular proteins. J.Mol.Biol.120,97-120.
\bibitem{gor3} Gibrat,J.F.,Garnier,J.,and Robson,B.1987.Further developments of protein secondary structure prediction using information theory.Newparameters and consideration of residue pairs,J.Mol,Biol.198,425-443.
\bibitem{pslt} Hobohm,U.,and Sander,C.1994.Enlarged representative set of protein structures.Protein Science 3,522-524.
\bibitem{dssp} Kabsch,W.,and Sander,C.1983.Dictionary of protein secondary structure:pattern recognition of hydrogen-boned and geometrical features.Biopolymers 22, 2577-2637.
\bibitem{kl} Kullback,S.,  Keegel,J.C.  and Kullback,J.H.  1959. Information Theory and Statistics. Wiley, New York.
\bibitem{kl1} Kullback,S. 1987. Topics in Statistical Information Theory, Springer, Berlin .
\bibitem{kl2} Sakamoto,T., Ishiguro,M.  and Kitagawa,G. 1986. Akaike Information Criterion Statistics, KTK Scientific, Tokyo.
\bibitem{scop} Murzin,A.G.,Brenner,S.E.,Hubbard,T.,and Chothia,C.1995.SCOP:a structural classification of proteins database for the investigation of sequences and structures.J.Mol.Biol.247,536-540.
\bibitem{rs} Rost,B.,and Sander,C.1993.Prediction of protein secondary structure at better than 70\% accuracy.J.Mol.Biol.232,584-599.
\bibitem{ssp} Solovyev,V.V.,and Salamov,A.A.1991.Method of calculation of discrete secondary structures in globular proteins.Mol.Biol.25(3),810-824.
\bibitem{ssp1} Solovyev,V.V.,and Salamov,A.A.1994.Predicting alpha-helix and beta-strand segments of globular protein.Comput.Appl.Biosci.10(6),661-669.
\bibitem{tayl} Taylor,W.R.1984.An algorithm to compare secondary structure predictions.J.Mol.Biol.173,512-521.
}
\end{thebibliography}
\end{document}